# The Role of Tunneling Oxide in the Low Frequency Noise of Multi-level Silicon Nitride ReRAMs

Nikolaos Vasileiadis
*Institute of Nanoscience and Nanotechnology*
*NCSR "Demokritos"*
Ag. Paraskevi, Greece
n.vasiliadis@inn.demokritos.gr

Alexandros Mavropoulis
*Institute of Nanoscience and Nanotechnology*
*NCSR "Demokritos"*
Ag. Paraskevi, Greece
a.mavropoulis@inn.demokritos.gr

Christoforos Theodorou
*Univ. Grenoble Alpes , Univ. Savoie Mont Blanc, CNRS*
*Grenoble INP, IMEP-LAHC*
Grenoble, France
christoforos.theodorou@grenoble-inp.fr

Panagiotis Dimitrakis
*Institute of Nanoscience and Nanotechnology*
*NCSR "Demokritos"*
Ag. Paraskevi, Greece
p.dimitrakis@inn.demokritos.gr

*Abstract*— This research explores the characteristics of two CMOS-compatible RRAM cells utilizing silicon nitride as the switching material. By employing SET/RESET pulse sequences, the study successfully attains four distinct and stable resistance states. To gain deeper insights, a Low-Frequency Noise (LFN) statistical analysis is conducted to investigate the role of a tunneling oxide between the bottom electrode and $SiN_x$ at various resistance levels. The findings from the LFN measurements strongly suggest that the multi-level high resistance switching primarily arises from variations in the number of nitrogen vacancies, which in turn modulate the conductivity of conductive filaments (CF). Notably, this modulation does not compromise the quality of the filament's surrounding interface. This research sheds light on the underlying mechanisms of RRAM cells and their potential for advanced memory applications.

*Keywords—Silicon Nitride; RRAM; Multi-Level; Noise;*

## I. INTRODUCTION

Resistive memories (RRAMs) offer a promising alternative to the scaling of nonvolatile memories [1]. RRAM crossbar arrays have applications in neuromorphic and in-memory computing hardware accelerators [1]. Recently, RRAM devices have been explored for use in quantum simulators as memristors to store qubits in digital quantum simulators [2]–[4]. The silicon nitride ($SiN_X$) dielectric is attractive due to its resistance to oxygen diffusion and humidity. The SiN-based RRAMs exhibit competitive resistance switching properties and $SiN_X$ is more resistant to oxygen and metal atoms diffusion [5], guaranteeing a better reliability. Functional SiN-based RRAM devices were recently demonstrated by our group for application in neuromorphic and in-memory computing [6]–[8], as well as TRNGs [9].

The present study focuses on the role of traps in resistive switching of MIS and MIOS memristive devices. More specifically, the $SiO_2$ layer between $SiN_x$ and Si substrate introduces a higher energy barrier reducing the leakage of trapped carriers in bulk $SiN_x$ to Si [10]. Also, the $SiO_2$/Si interface has a better quality than the $SiN_x$/Si interface. Thus, the carrier exchange (trapping/de-trapping) between Si and $SiN_x$ through boarder traps is minimized [11].

A LFN statistical analysis is conducted to investigate the role of a tunneling oxide between the bottom electrode and $SiN_x$ at various conductance levels. The findings from the LFN measurements strongly suggest that the multi-level high resistance switching primarily arises from variations in the number of nitrogen vacancies, which in turn modulate the conductivity of conductive filaments (CF).

Section II provides the experimental details such as fabrication and measurement setup. Section III provides the results and the discussion.

## II. DEVICES AND MEASUREMENT SET-UP

The examined Metal-Insulator-Silicon (MIS) and Metal-Insulator-Oxide-Silicon (MIOS) structures were fabricated according to the detailed procedures outlined in [3]. To ensure precise characterization, cross-section Transmission Electron Microscopy (XTEM) measurements were performed on both samples, and the results are presented in Fig. 1. Notably, the silicon nitride ($SiN_x$) layers exhibited consistent thicknesses of (6.7 ± 0.3) nm and (6.2 ± 0.3) nm for MIS and MIOS configurations, respectively. These nitride layers displayed an amorphous nature with negligible surface and interface roughness, ensuring homogeneity throughout the samples. The bottom electrode consisted of a heavily doped silicon substrate ($\rho$=4 mΩ·cm), while the top electrode was a 30nm PVD Cu layer.

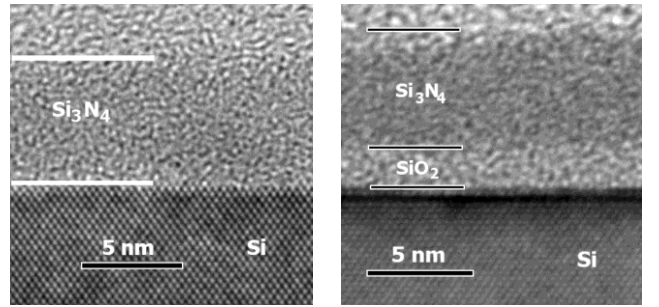

Fig. 1. XTEM micro-images for structures (a) without and (b) with tunneling oxide. The oxide layer thickness was measured (2.1 ± 0.3) nm.

For programming and investigating multiple resistance levels, a custom experimental setup was designed [12]. Achieving multi-level resistance required precise control of compliance current and a flexible pulse tuning protocol. To fulfill this requirement, a custom setup was constructed, as depicted in Fig. 2(a). In this setup, a DAQ (Data Acquisition) card, specifically the NI-PCIO-MIO-16E, was connected to an SR570 I/V converter through a low-noise junction box (NI BNC 2110). Further details regarding the experimental setup are available in [12].

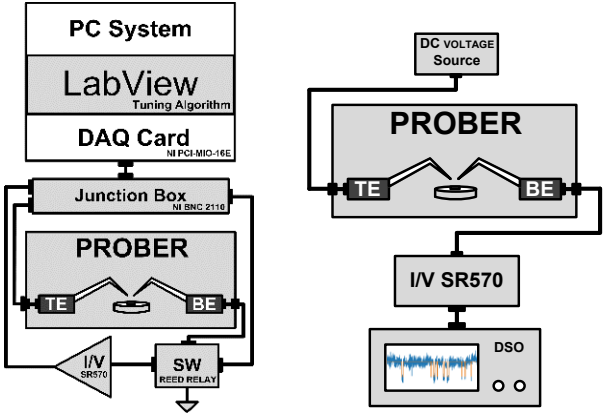

Fig. 2. Block diagram of the experimental multi-level resistance tuning setup (left) and Random telegraph noise measurement setup (right), adopted from [5].

Figure 2(b) illustrates the experimental arrangement employed for Random Telegraph Noise (RTN) measurements. Following the establishment of the memristors at predefined resistance levels, RTN data were recorded using a digital oscilloscope, the Agilent 7000 series Oscilloscope (DSO), over a duration of 100 seconds with a sampling interval of 25 μs. This recording was facilitated through an I/V converter. It is noteworthy that all experiments were conducted at room temperature, utilizing memristors with a uniform area of 100 μm × 100 μm.

### III. EXPERIMENTAL RESULTS AND DISCUSSION

#### A. MIS devices

The Fig. 3 presentation encompasses some examples of time-domain drain current noise measurements conducted on Device #1 from the MIS and #6 from the MIOS ensemble. These measurements span a spectrum of set resistances, providing a comprehensive view of the device behavior under varying conditions. Notably, the presence of a discernible RTN signal becomes pronounced at the lowest resistance level, signifying a heightened sensitivity to noise phenomena in this particular regime.

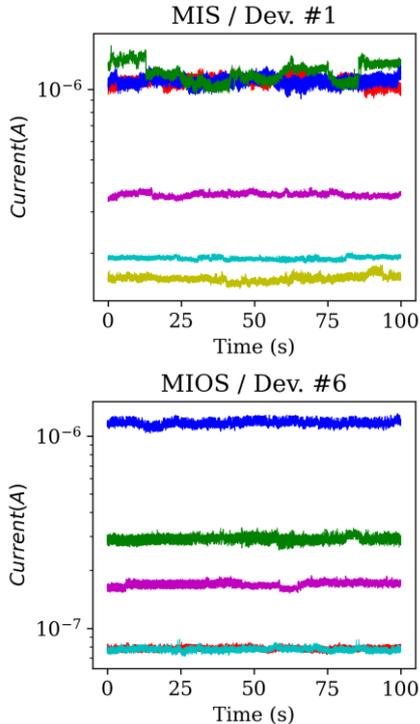

Fig. 3. Example of measured time domain current signals at 0.1V for various levels of the programmed resistance of a MIS (top) and a MIOS (bottom) device: RTN signal appears at some resistance levels.

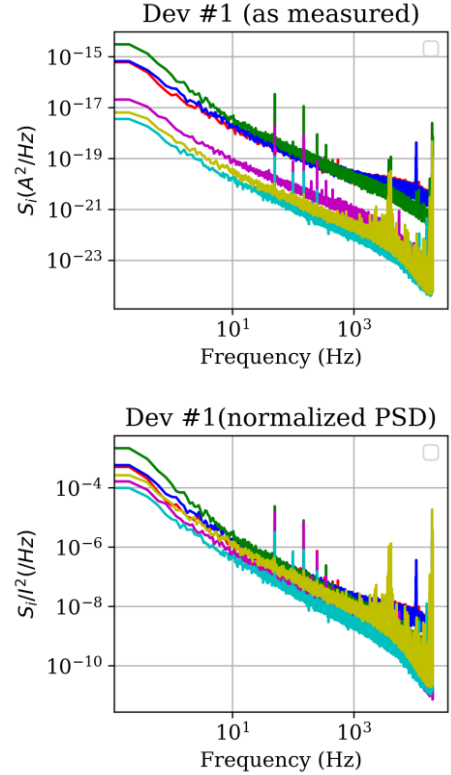

Fig. 4. Measured current PSD, $S_i$, of Fig. 2 signals, before (left) and after (right) normalization with current for a MIS device.

In order to gain further insight into the intricacies of the measured noise levels, we turn our attention to Fig. 4 (top), where the corresponding noise spectra, denoted as $S_i$, are meticulously displayed. What becomes strikingly evident is the considerable dispersion in the measured noise levels, spanning almost two orders of magnitude. This high degree of variability underscores the complexity and multifaceted nature of the underlying noise phenomena. It highlights the need for an in-depth analysis to comprehend the various contributing factors within the MIS ensemble effectively.

However, an intriguing observation arises when examining the normalized $S_i/I^2$, as depicted in Fig. 4 (bottom). Here, we note that the resistance-to-resistance variability is significantly reduced, amounting to less than one decade. This finding indicates that the LFN magnitude closely tracks changes in resistance, a relationship established through proportionality to $I^2$. It is important to emphasize that the bias voltage remained constant at 0.1V during RTN measurements. Therefore, the fluctuations in current, as elucidated in Fig. 3 and 4, can be attributed to variations in the device resistance. This variation, characterized by multi-level resistance, could arise from either the creation of new Conductive Filaments (CFs) or the modulation of existing ones.

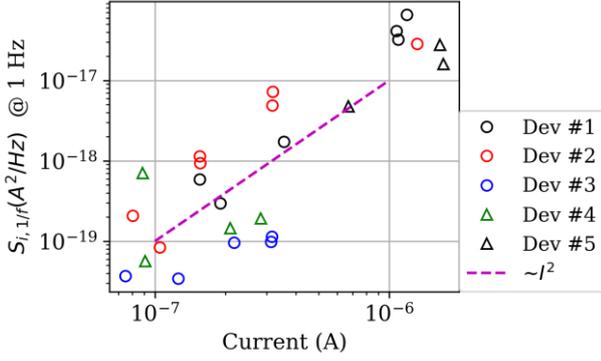

Fig. 5. Extrapolated *1/f* noise component amplitude at *f* = 1 Hz for different programming resistances versus the measured DC current for all the examined MIS devices.

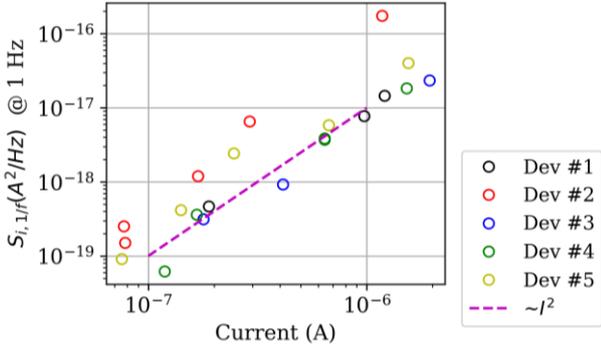

Fig. 6. Extrapolated *1/f* noise component amplitude at *f* = 1 Hz versus measured DC current for all the examined MIOS devices.

Finally, our investigation extends to Fig. 5, where the *1/f* component of each spectrum in Fig. 4 is extrapolated at 1 Hz and plotted against the measured DC current. This presentation reveals a distinct $\sim I^2$ trend, particularly prominent in devices exhibiting substantial resistance changes, surpassing 2-3 times. However, it's worth noting that some devices exhibit higher noise levels than others, a common phenomenon in nano-scale devices where defect density varies significantly from device-to-device, as reported in previous studies [13].

These intriguing findings offer valuable insights into the complex interplay of noise sources and device characteristics within the MIS ensemble, opening avenues for further exploration and refinement in nanoscale electronic systems.

### B. MIOS devices

The identical measurement procedure and analytical methodology were meticulously adhered to in the case of MIOS devices. In Fig. 6, we present a graphical representation of the *1/f* component of the spectral data at 1 Hz versus the DC current at various resistance levels. It is noteworthy that all of the examined devices exhibit a harmonious alignment in terms of noise levels concerning the identical current values, and they distinctly adhere to a discernible quadratic relationship with the DC current, approximately proportional to $I^2$. Only one device (Dev.#2) declines from this trend exhibiting a stronger power-law dependence than $I^2$, which seems to be an outlier. The number of measured devices is small in order to reach solid conclusions, nevertheless suggests that MIOS devices exhibit a notably superior degree of control over variability concerning defect density contributing to LFN. This notable control could potentially be ascribed to the presence of a tunneling oxide layer within these devices. Although the same conclusion has been investigated by c-AFM measurements [14], further investigation is needed.

### C. LFN comparison between MIS and MIOS SiN RRAMs

Given that the LFN characteristics of the majority of devices, encompassing both MIS and MIOS structures, exhibit a linear relationship with the square of DC current, we proceeded to calculate the logarithmic mean of the normalized PSD, denoted as $S_i/I^2$, across all devices at various resistance levels. The outcome of this normalization procedure is thoughtfully displayed in Fig. 7, where they are juxtaposed with the average noise level derived from a set of five devices.

Intriguingly, the average normalized noise levels for both MIS and MIOS RRAMs harmoniously converge, despite the presence of a non-negligible variability factor. It should be emphasized that the average (Fig. 7) of the log-mean values for MIS and MIOS are almost the same.

### D. Interpretation of LFN behavior

The stability of $S_i/I^2$ with $R$ could be explained considering the fluctuation and CF formation mechanisms. So, assuming a *1/f* carrier number fluctuations model for a cylindrical metal-like CF, the following relation [15] should be true:

$$\frac{S_i}{I^2} = \frac{S_N}{N^2} = \frac{kTN_{it}2\pi r t_{SiN}}{n^2(\pi r^2 t_{SiN})^2 f} = \frac{2kTN_{it}}{n^2\pi r^3 t_{SiN} f} \quad (1)$$

where $r$ the filament radius, $t_{SiN}$ its height, $n$ the equivalent free carrier density and $N_{it}$ the surface trap density (1/cm$^2$/eV) regarding the traps on the surface surrounding the filament.

Equation (1) suggests that if the creation of new CFs modify the resistance value, then $S_i/I^2$ would be reciprocally proportional to their number. This is because increasing the number of CFs the number of $N_{it}$ is increasing in proportion. Similarly, if the resistance is modified due to change in the radius of the CF, then $S_I/I^2$ would change as $\sim 1/r^3$. In addition, if the multi-resistance levels are due to the change of $n$, then the normalized PSD should change accordingly, i.e., $\sim 1/n^2$. Also, in [16] a constant $\Delta R/R$ was noticed for high resistance RRAM values where the filament has a semiconductor-like behavior with no screening effect for the filament's surrounding traps. Therefore, according to the LFN measurements, it is safe to assume that the multi-level high resistance switching is mainly attributed to the change in the number of nitrogen vacancies that modulate the CF conductivity, and without affecting the quality of the filament's surrounding interface.

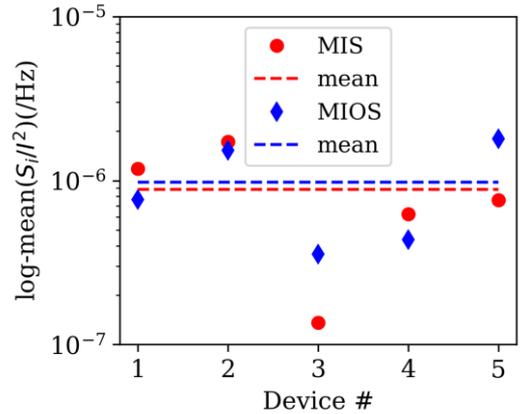

Fig. 7. Mean values of normalized PSD for all resistance levels for each device, along with their average (dashed lines) across all the examined MIS and MIOS devices.

## IV. Conclusion

The utilization of LFN measurements in multi-level MIS and MIOS RRAM cells, primarily constructed using $SiN_x$, has offered valuable insights. The experimental results unveiled the favorable impact of the tunneling oxide at the $SiN_x$/Si interface, which seems to improve the switching performance considering that setting the devices to lower resistances was easier, while keeping the noise at the same levels. Additionally, some light was shed on the underlying nature of the resistance change mechanism and the lore of the nitrogen vacancies within the conductive filament.